\documentclass[12pt,twoside]{article}
\usepackage{fleqn,espcrc1}
\usepackage[dvips]{graphicx}
\usepackage{color} 
\newcommand{\xp}{\ensuremath{x_{I\!\!P}}}

\newcommand{\FD}{\ensuremath{F_{2}^{D(3)}}}
\newcommand{\dcs}{\ensuremath{\sigma_{d}}}
\newcommand{\aem}{\ensuremath{\alpha_{em}}}

\newcommand{\D}{\ensuremath{\mbox{d}}}

\begin{document}
\begin{flushright}
\end{flushright}
{\large  Predicting $F_{2}^{D(3)}$ from the colour glass condensate model} \\

\noindent J.~R.~Forshaw$^{a}$,
R.~Sandapen$^{b}$
and G.~Shaw$^{a}$

\vspace*{0.5cm}

\noindent $^{a}$Theoretical Physics Group, Department of Physics and Astronomy,\\
The University of Manchester, M13 9PL, UK
\vspace*{0.5cm}

\noindent $^{b}$Theoretical Physics Group, Department of Physics, Engineering Physics and Optics\\
Laval University, G1K 7P4, Quebec, Canada.
\vspace*{0.5cm}

\begin{abstract}
We confront the colour glass condensate motivated dipole model parameterization
of Iancu, Itakura and Munier with the available HERA data on the diffractive
structure function $\FD$ and with existing dipole model parameterizations. Reasonably good
agreement is found with only two adjustable parameters. We caution against interpreting
the success of the model  as compelling evidence for low-$x$ perturbative saturation dynamics.

\end{abstract}
\section{Introduction}

Some years ago, it was shown~\cite{bib:GW2,bib:FKS2}  that two
phenomenological colour dipole models 
- the saturation model of Golec-Biernat and W\"usthoff~\cite{bib:GW1} and the
two-component model of Forshaw, Kerley and Shaw ~\cite{bib:our} - yielded
 a rather good description  of the
 diffractive deep inelastic scattering (DDIS) data~\cite{bib:H1,bib:ZEUS}. 
These successes were achieved
without adjusting any of the parameters
of the models, which had been previously determined by fits to   the deep
inelastic structure function data at  small-$x$  ~\cite{bib:GW1,bib:our}.
In both cases the DDIS structure function  $F_{2}^{D(3)}(\beta,x,Q^2)$ was
found to be dominated by  the $q \bar{q}$ dipole contribution at large
$\beta$, corresponding to diffractively produced states with invariant 
mass $M_X^2 \ll Q^2$, but  at small $\beta$, corresponding to
$M_X^2 \gg Q^2$, a  $q \bar{q} g$ contribution becomes important (see
below). Subsequently, both models have been successfully applied to a variety
of other processes \cite{bib:mss02,bib:fm03,bib:fss1,bib:fss2,bib:cs01}.

More recently a new colour dipole model, the colour glass condensate (CGC)
model of  Iancu, Itakura and Munier \cite{bib:iim03}, has aroused considerable
interest. This model  can be thought of
as a development of the Golec-Biernat--W\"usthoff saturation model. 
However, while still largely
a phenomenological parameterisation, the authors  claim that it contains
the main features of the ``color glass condensate'' regime, where the 
gluon densities are high and non-linear effects become important.
The parameters of the model have again been fixed by fitting the structure
function data, which are now extremely precise, and the model has subsequently
been shown to yield a good description of both $\rho$, $\phi$ \cite{bib:fss1}
and $J/\Psi$ \cite{bib:fss2} electroproduction data, with reasonable choices
for the vector meson wavefunctions. However it has not yet been applied to
DDIS, an omission which we propose to rectify in this short paper. 

\section{The CGC model for $F_2(x,Q^2$)}

\begin{figure}[htb]
\begin{center}
\includegraphics[width=10cm, height=3cm]{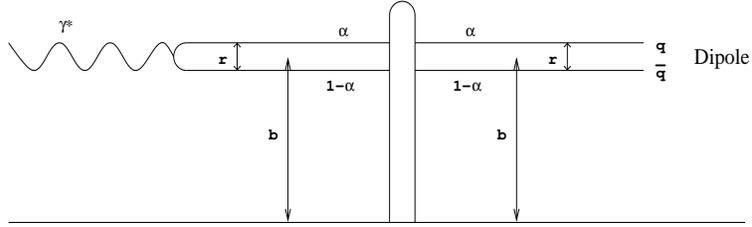}
\caption{The dipole model for $F_2(x,Q^2)$.}
\label{fig:fluct}
\end{center}
\end{figure}

In the colour dipole model (Fig.~\ref{fig:fluct}), the dipole cross-section
$\dcs(s^*,r,\alpha)$ 
is related to the photon-proton cross-section via 
\begin{equation}
\sigma^{L,T}_{\gamma^{*}p} = \int d\alpha \;  d^2 r \ 
|\psi_{L,T}(\alpha,r)|^{2} \dcs(s^*,r,\alpha)  
\end{equation} 
where $r$ is the transverse separation of the q\={q} pair, 
$\alpha$ is the fraction of the incoming photon light-cone energy 
carried  by the quark, and the variable $s^*$ is chosen to be either 
 $s = W^2$, the squared CMS energy of the 
photon-proton system, or the Bjorken scaling variable $x$. Henceforth we assume
that the dipole cross-section is independent of $\alpha$, i.e. we write $\sigma_d(s^*,r)$.

In the CGC model, $s^*=x$ and the  longitudinal and
transverse components of the light-cone photon wave function  are
assumed to be given by the tree level QED expressions~\cite{bib:dcs_nik,bib:wf}:
\begin{eqnarray}
  \label{eq:psi^2}
  |\psi_{L}(\alpha,r)|^{2} & =  & \frac{6}{\pi^{2}}\aem\sum_f e_{f}^
{2}Q^{2}\alpha^{2}(1-\alpha)^{2} K_{0}^{2}(\epsilon r) \\
  |\psi_{T}(\alpha,r)|^{2} & = & \frac{3}{2 \pi^{2}}\aem\sum_{f} e_{f}^
{2} \left\{[\alpha^{2} + (1-\alpha)^{2}] \epsilon^{2} K_{1}^{2}(\epsilon r) + m_{f}^{2} 
K_{0}^{2}(\epsilon r) \right\} 
\end{eqnarray}
where 
\(
 \epsilon^{2} = \alpha(1-\alpha)Q^{2} + m_{f}^{2}\; ,
\) 
$K_{0}$ and $K_{1}$ are modified Bessel functions and the sum is over 
quark flavours $f$ with quark masses $ m_f$.  The CGC dipole cross-section
is assumed to be of the form
\begin{eqnarray}
\sigma_d(x,r) &=& 2 \pi R^2 {\cal N}_0 \left( \frac{r Q_s}{2} 
\right)^{2\left[\gamma_s + \frac{\ln(2/rQ_s)}{\kappa \lambda \ln(1/x)}\right]} 
\hspace*{1cm} \mathrm{for} \hspace*{1cm} rQ_s \le 2 \nonumber \\
&=& 2 \pi R^2 \{1 - \exp[-a \ln^2(brQ_s)]\} \hspace*{1cm} \mathrm{for} 
\hspace*{1cm} rQ_s > 2~,
\label{cgc-dipole}  
\end{eqnarray} 
where the saturation scale $Q_s \equiv (x_0/x)^{\lambda/2}$ GeV. The
coefficients $a$ and $b$ are uniquely determined by ensuring continuity of
the cross-section and its first derivative at $rQ_s=2$. The leading order
BFKL equation fixes $\gamma_s = 0.63$ and $\kappa = 9.9$. The coefficient
${\cal N}_0$ is strongly correlated to the definition of the saturation
scale and the authors find that the quality of fit to $F_{2}$ data is only weakly dependent
upon its value. For a fixed value of ${\cal N}_0$, there are therefore
three parameters which need to be fixed by a fit to the structure function data, i.e.
$x_0$, $\lambda$ and $R$. 
In this paper, we take $N_0 = 0.7$ and a light quark mass of $m_q = 140$ MeV,
for which the fit values are 
$x_0 = 2.67 \times 10^{-5}$, $\lambda = 0.253$ and $R = 0.641$ fm. We take 
$x = Q^2/(Q^2+W^2)\times (1+4m_q^2/Q^2)$\footnote{For the $Q^2$ values we consider in this paper,
the mass correction is unimportant.}
and for the charm quark contribution we take $m_c = 1.4$~GeV.

\section{The CGC model for $\FD $}
\begin{figure}[htb]
\begin{center}
\includegraphics[height=6cm,width=6cm]{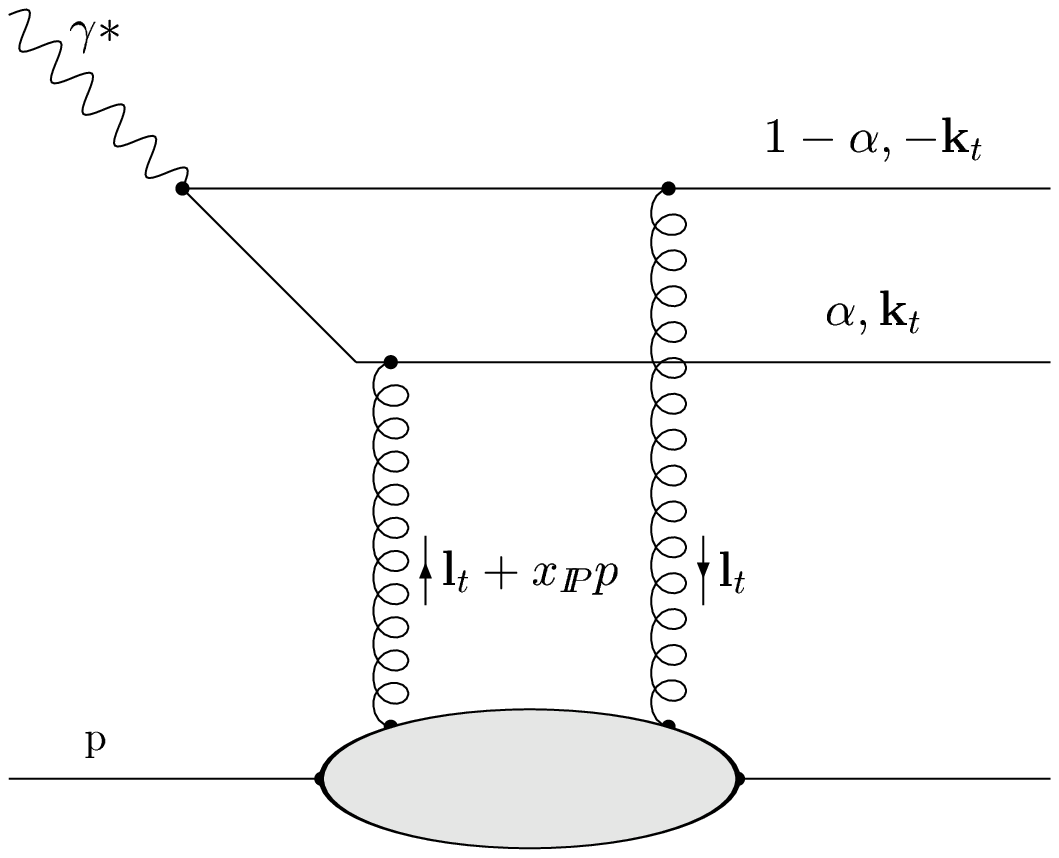}
\includegraphics[height=6cm,width=6cm]{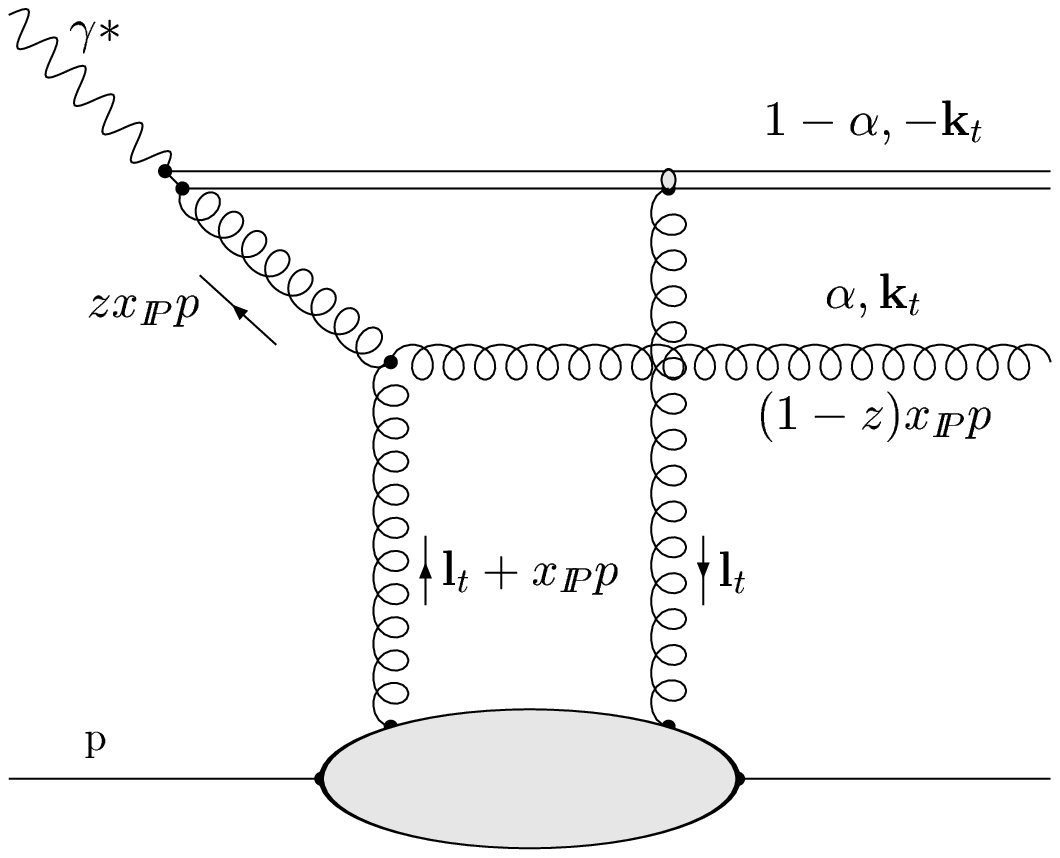}
\caption{The q\={q} and q\={q}g contributions to $ \FD$.}
\label{fig:feynman}
\end{center}
\end{figure}

To calculate the contribution of the quark-antiquark dipole  to $ \FD$ we made
use of expressions  derived from a momentum space treatment. We calculate the 
contribution of the higher q\={q}g Fock state using an effective two-gluon dipole 
description \cite{bib:GW2,bib:mw.der}. Typical Feynman diagrams are shown in
Figure~\ref{fig:feynman}. 

Defining
\begin{equation}
  \Phi_{0,1}  \equiv  \left( 
\int_{0}^{\infty}r \D r K_{0 ,1}(\epsilon r)\sigma_{d}(\xp,r) J_{0 ,1}(kr) \right)^2,
\end{equation}
we have for the longitudinal and transverse q\={q} components
\begin{equation}
  x_{I\!\!P}F^{D}_{q\bar{q},L}(Q^{2}, \beta, x_{I\!\!P})=
\frac{3 Q^{6}}{32 \pi^{4} \beta b}\cdot \sum_{f} e_{f}^{2} 
\cdot 2\int_{\alpha_{0}}^{1/2} \D \alpha \alpha^{3}(1-\alpha)^{3} \Phi_{0},
\end{equation}
\begin{equation}
 x_{I\!\!P}F^{D}_{q\bar{q},T}(Q^{2}, \beta, x_{I\!\!P}) =  
     \frac{3 Q^{4}}{128\pi^{4} \beta b} \cdot \sum_{f} e_{f}^{2} \cdot
 2\int_{\alpha_{0}}^{1/2} \D \alpha \alpha(1-\alpha) 
\left\{ \epsilon^{2}[\alpha^{2} + (1-\alpha)^{2}] \Phi_{1} + m_f^{2} \Phi_{0}  \right\}   
\end{equation}
where the lower limit of the integral over $\alpha$ is given by
\(
\alpha_{0} = (1/2)\left(1 - \sqrt{1 - 4m_{f}^{2}/M_{X}^{2}}\right)
\)
and $b$ is the slope parameter which, unless otherwise stated, we take to be 7.2~GeV$^{-2}$
\cite{bib:slope}. 

For the q\={q}g term we have\footnote{Following~\cite{bib:FKS2}, we
 have inserted a missing factor of 1/2 compared with the expression in~\cite{bib:GW2}.}
 \begin{eqnarray}
   \lefteqn{x_{I\!\!P}F^{D}_{q\bar{q}g}(Q^{2}, \beta, x_{I\!\!P}) 
  =  \frac{81 \beta \alpha_{S} }{512 \pi^{5} b} \sum_{f} e_{f}^{2} 
 \int_{\beta}^{1}\frac{\mbox{d}z}{(1 - z)^{3}} 
 \left[ \left(1- \frac{\beta}{z}\right)^{2} +  \left(\frac{\beta}{z}\right)^{2} \right] } \\
  & \times & \int_{0}^{(1-z)Q^{2}}\mbox{d} k_{t}^{2} \ln \left(\frac{(1-z)Q^{2}}{k_{t}^{2}}\right) 
   \left[ \int_{0}^{\infty} u \mbox{d}u \; \sigma_{d}(u / k_{t}, x_{\xp}) 
   K_{2}\left( \sqrt{\frac{z}{1-z} u^{2}}\right)  J_{2}(u) \right]^{2}.
\end{eqnarray} 
The normalization of this component is rather uncertain; unless otherwise stated we
take $\alpha_{S} = 0.2$.

\begin{figure}[htb]
\hspace*{-2cm}
\includegraphics[width=18cm]{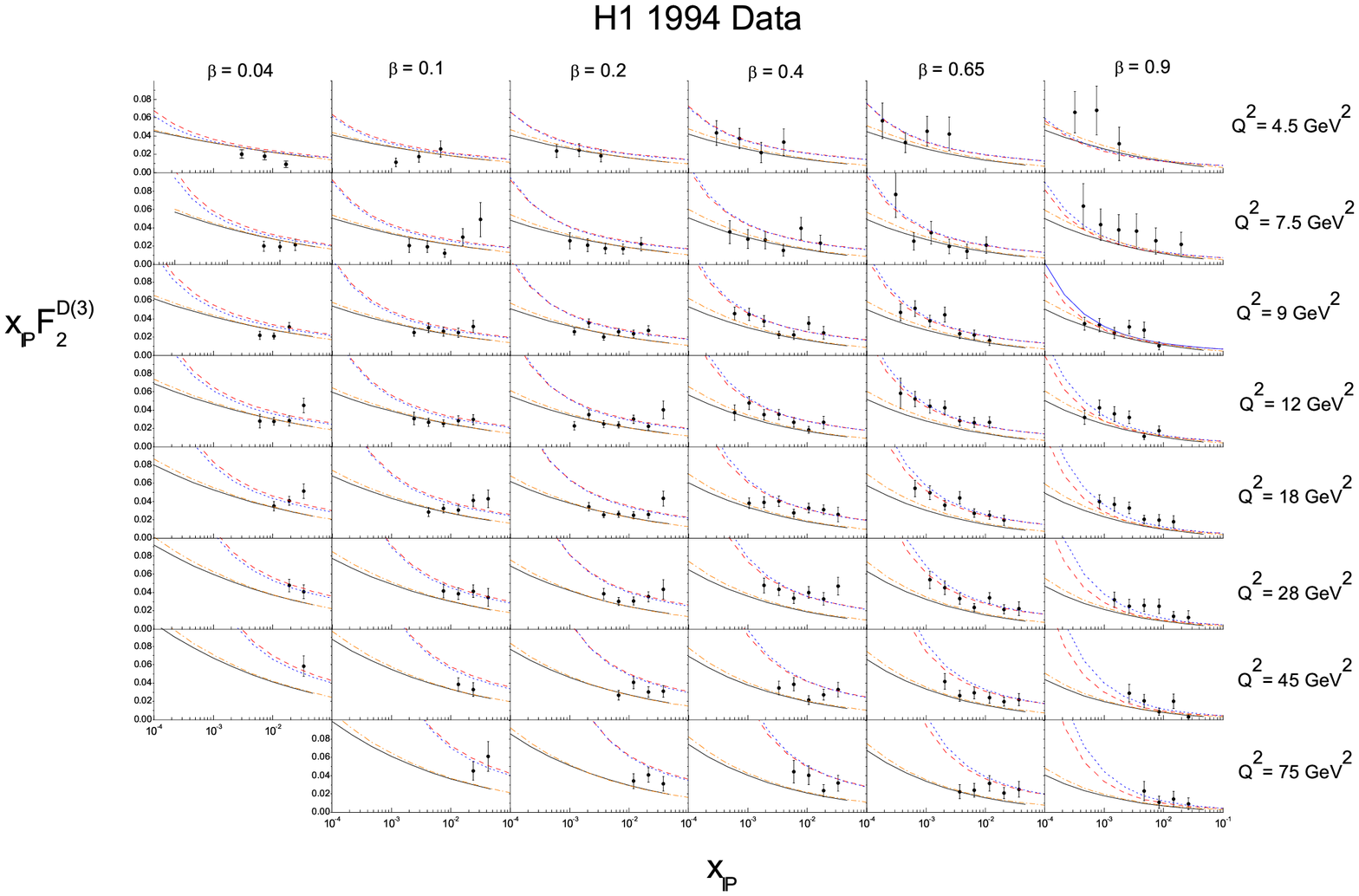}  
  \caption{Predictions for $\xp \FD$ compared with H1 1994 data. 
Solid black line: CGC model;
Dashed red line: FKS (2002); Dotted blue line: FKS (1999); Dash-dot orange line: GW model.   
}
  \label{fig:f2d3h1}
\end{figure}

\begin{figure}[htb]
\hspace*{-1cm}
\includegraphics[width=14cm]{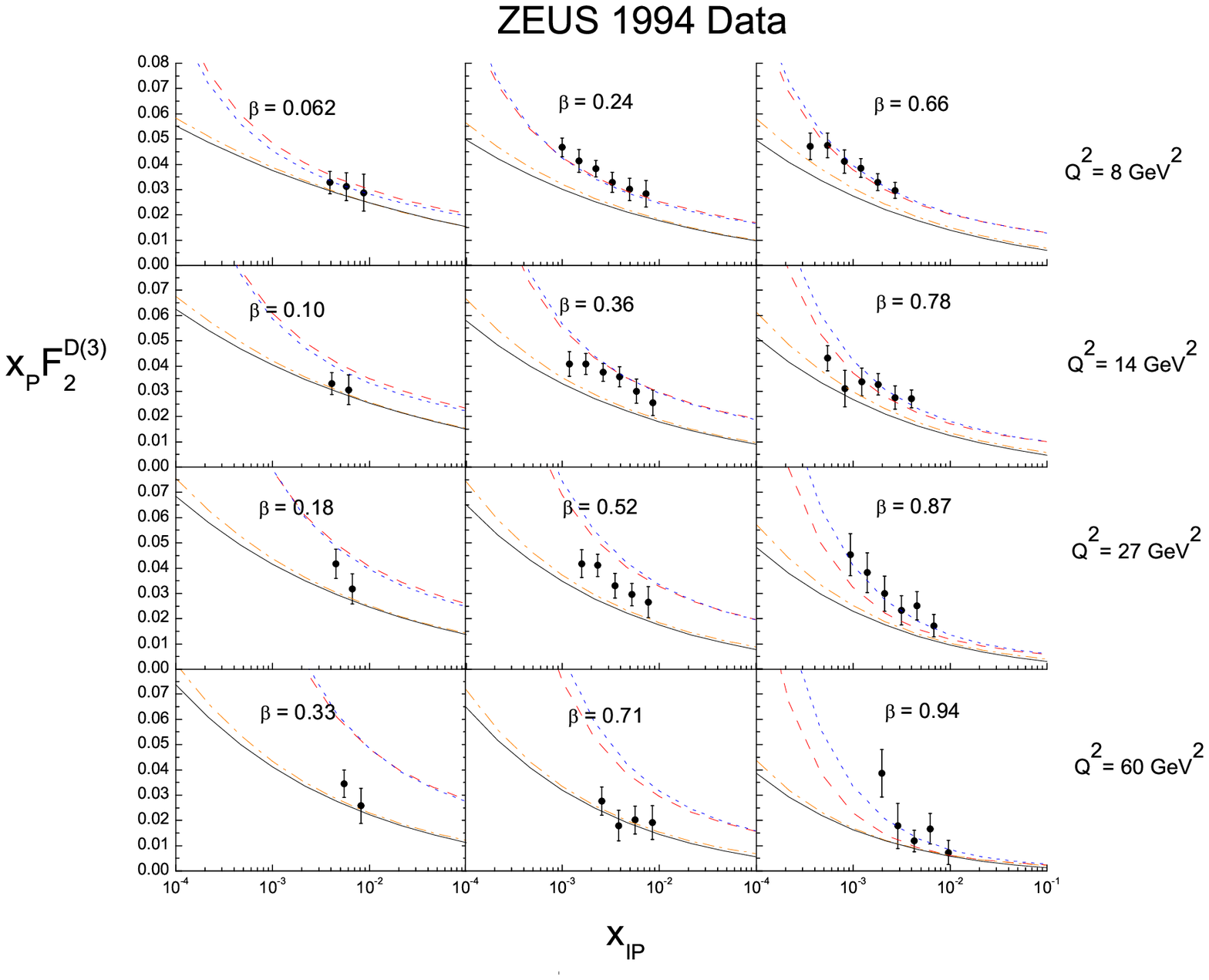}  
  \caption{Predictions for $\xp \FD$ compared with ZEUS 1994 data. 
Solid black line: CGC model;
Dashed red line: FKS (2002); Dotted blue line: FKS (1999); Dash-dot orange line: GW model.   
}
\label{fig:f2d3zeus} 
\end{figure}

Plots of the contributions to $\xp \FD$ calculated from these expressions are
compared with H1 1994 data~\cite{bib:H1} in Figure~\ref{fig:f2d3h1}
and with the ZEUS 1994 data~\cite{bib:ZEUS} in Figure~\ref{fig:f2d3zeus}. The CGC model predictions
are shown as the solid black curves in each plot. Also shown for comparison are the predictions of
the Golec-Biernat \& W\"usthoff (GW) saturation model \cite{bib:GW2} and the predictions of the FKS
model \cite{bib:our}. For the GW model, the curves are exactly as in \cite{bib:GW2} except that we use a
diffractive slope $b=7.2$~GeV$^{-2}$ rather than the GW choice of $b=6$~GeV$^{-2}$. For the FKS
model we show two curves: the dashed red line uses the more compact parameterization of the dipole
cross-section presented first in \cite{bib:mss02} whilst the blue dotted line is generated using
the original ``Fit 1'' parameterization of \cite{bib:our}. As anticipated there is not much difference
between the two FKS curves. There is also not very much difference between the CGC and GW curves, 
reflecting the fact that these two models use quite similar dipole cross-sections.  

\begin{figure}[htb]
\hspace*{-2cm}
\includegraphics[width=18cm]{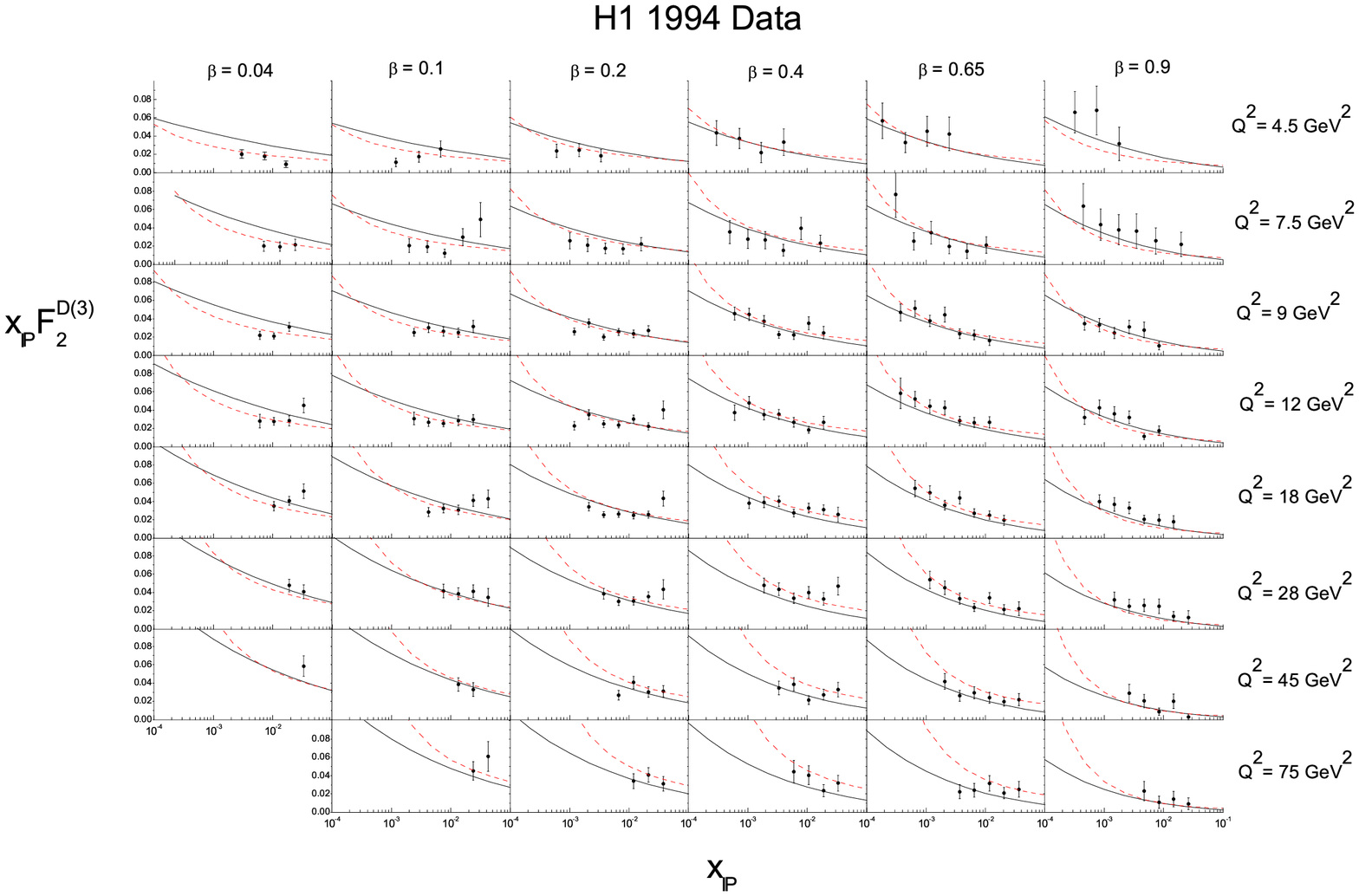}  
\caption{Predictions for $\xp \FD$ compared with H1 1994 data after tuning the $b$ parameter and
$\alpha_s$ (see text).  
Solid black line: CGC model; Dashed red line: FKS (2002).
}
\label{fig:f2d3modify} 
\end{figure}

All our predictions contain no adjustable parameters in the dipole cross-section itself. However, we
are free to adjust the forward slope for inclusive diffraction, $b$, within the range acceptable to
experiment. This simply affects the overall normalization of $\FD$. One can substantially improve the
quality of the CGC fit to the data by lowering $b$ towards the lower end acceptable to experiment,
i.e. $b = 5.5$~GeV$^{-2}$. We are also somewhat free to vary the value of $\alpha_s$ used
to define the normalization of the $q\bar{q}g$ component, which enters at low to intermediate values
of $\beta$, indeed choosing $\alpha_s = 0.15$ for the FKS model leads to a much improved fit. In
Figure~\ref{fig:f2d3modify} we compare the CGC model, with the lower $b$ parameter, 
and the FKS model, with the lower value of $\alpha_s$, 
to the H1 data. Both models now agree rather well with the data.
   
\begin{figure}[htb]
\hspace*{-1cm}
\includegraphics[width=14cm]{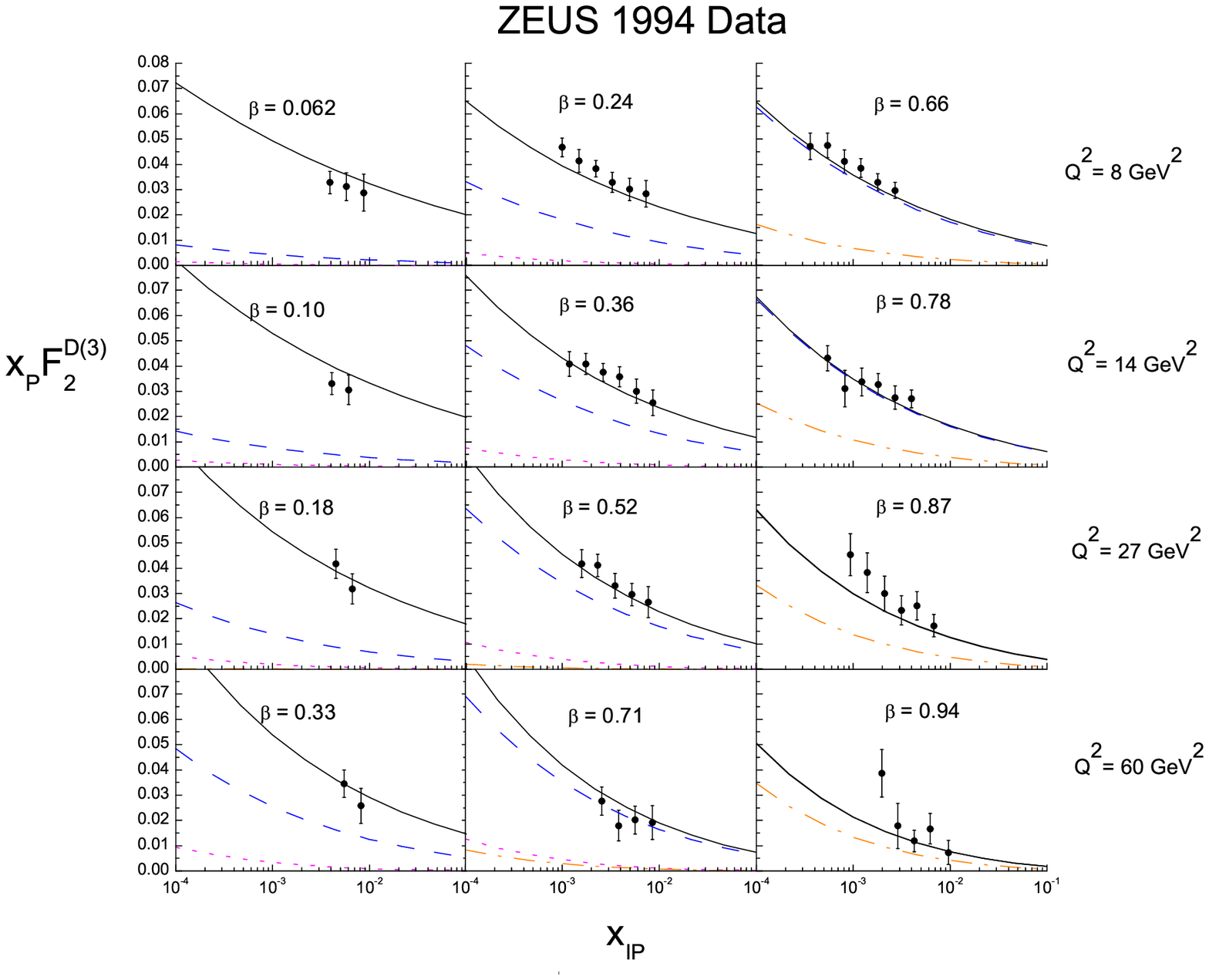}  
\caption{Various contributions to $\xp \FD$ compared with ZEUS 1994 data using the CGC dipole 
model. 
Solid black line: Total contribution; Dashed blue line: $q \bar{q}$ contribution; Dotted red line: 
$c \bar{c}$ contribution; Dashed-dotted orange line: longitudinal $q\bar{q}$ contribution. 
}
\label{fig:f2d3components} 
\end{figure}

Finally, in Figure \ref{fig:f2d3components} we show a breakdown of the CGC model, with 
the lower value of $b=5.5$~GeV$^{-2}$, into its various components and its comparison with the
ZEUS data. The solid black line is again the total contribution whilst 
the blue dashed line is the
contribution from $q \bar{q}$ dipoles (light quarks only), the red dotted line is the 
contribution from $c \bar{c}$ dipoles and orange dash-dot line is the contribution from the light
quark dipoles produced by longitudinally polarized photons. We note that, in the region where
the $q \bar{q}$ contribution is dominant (i.e. at larger values of $\beta$), approaches based 
upon the dipole model have very little room for manoeuvre. In particular, only the 
normalization is uncertain within the range allowed by the error on the measurement of the $b$ 
parameter.

\section{Conclusions}
We have used the dipole parameterization of 
Iancu, Itakura and Munier \cite{bib:iim03} to
predict the diffractive structure function $\FD$. This parameterization is anticipated to capture
some of the essential dynamics of the colour glass condensate approach, in which
saturation in $x$ at fixed $Q^2$ is an essential feature. Agreement with the data is
reasonably good provided that one accepts a forward slope for diffraction of $5.5$~GeV$^{-2}$.
We note that the CGC predictions are very similar to the previous predictions of Golec-Biernat \&
W\"usthoff \cite{bib:GW2}. In a previous paper, we have shown that the CGC model is also capable of 
describing the data on exclusive vector meson production \cite{bib:fss1}. 

However, we stress 
that the same data are also consistent with a ``two pomeron'' model \cite{bib:FKS2,bib:our,bib:DL} in
which there is no low $x$ saturation. Indeed Figure \ref{fig:f2d3modify} of this paper
compares the predictions of the FKS and CGC models to the $\FD$ data. As such we conclude that the 
data are not yet precise enough, nor do they extend to sufficiently small values of $\xp$, to 
descriminate between these very different theoretical approaches.

\section{Acknowledgement}
This research was supported in part by a UK Particle Physics and Astronomy 
Research Council grant number PPA/G/0/2002/00471.
\end{document}